\documentclass{emulateapj}
\shorttitle{Exploring the MGC Color-Structure Bimodality}
\shortauthors{Cameron et al.}

\begin{document}
\title{The Millennium Galaxy Catalogue: Exploring the Color-Concentration Bimodality via Bulge-Disc Decomposition}

\author{Ewan Cameron\altaffilmark{1,2}, Simon P. Driver\altaffilmark{1}, Alister W. Graham\altaffilmark{3}, and Jochen Liske\altaffilmark{4}}

\altaffiltext{1}{SUPA (Scottish Universities Physics Alliance), School of Physics and Astronomy, University of St Andrews, North Haugh, St Andrews, KY16 9SS, Scotland; ec60@st-andrews.ac.uk, spd3@st-andrews.ac.uk}

\altaffiltext{2}{Research School of Astronomy and Astrophysics, Mount Stromlo Observatory, Cotter Road, Weston Creek, A.C.T., 2611, Australia}

\altaffiltext{3}{Centre for Astrophysics and Supercomputing, Swinburne University of Technology, Hawthorn VIC 3122, Australia; agraham@astro.swin.edu.au}

\altaffiltext{4}{European Southern Observatory, 85748 Garching, Germany; jliske@eso.org}

\begin{abstract}
We investigate the origin of the galaxy color-concentration bimodality at the bright-end of the luminosity function ($M_B - 5 \log h_{70} < -18$ mag) with regard to the bulge-disc nature of galaxies.  Via (2D) surface brightness profile modeling with GIM2D, we subdivide the local galaxy population in the Millennium Galaxy Catalogue into one-component and two-component systems.  We reveal that one-component (elliptical and disc-only) systems define the two peaks of the galaxy color-concentration distribution (with total stellar mass densities of $0.7 \pm 0.1$ and $1.3 \pm 0.1$ $\times$ $10^8$ $h_{70}$ $M_\odot$ Mpc$^{-3}$ respectively), while two-component systems contribute to both a bridging population and the red, concentrated peak (with total stellar mass densities of $1.1 \pm 0.1$ and $1.8 \pm 0.2$ $\times$ $10^8$ $h_{70}$ $M_\odot$ Mpc$^{-3}$ respectively).  Moreover, luminous, `bulge-less, red discs' and `disc-less, blue bulges' (blue ellipticals) are exceptionally rare (with \textit{volume}-densities of $1.7 \pm 0.3$ and $1.1 \pm 0.1$ $\times$ $10^{-4}$ $h^{3}_{70}$ Mpc$^{-3}$ respectively).  Finally, within the two-component population we confirm a previously-reported correlation between bulge and disc color (with a mean offset of only $\left<(u-r)_{\mathrm{bulge}}-(u-r)_{\mathrm{disc}}\right> = 0.22 \pm 0.02$ mag).\\
\end{abstract}


\section{Introduction}\label{intro}
The rigorous, statistical confirmation of bimodality within the luminous ($M_B - 5 \log h_{70} < -18$ mag), nearby galaxy population has proven a key result to emerge from modern wide-field imaging surveys---implying the existence of two principal evolutionary pathways for bright systems.   This bimodality is clearly evident in the distribution of local galaxies in various \textit{global} (i.e., total) observable pairs, including color-magnitude \citep{str01,bal04}, concentration (log $n$)-magnitude\footnote{The S\'ersic index is monotonically related to the degree of central concentration within a galaxy \citep{tru01} and is used here as a proxy for concentration.  In this paper the symbol $n$ is used exclusively to represent the \textit{global} S\'ersic index, i.e., the inverse-exponent in the best-fitting, single S\'ersic function to the total galaxy light profile.  In the context of our two-component model, where only the central excess above an outer exponential (i.e., the bulge) is fit with a S\'ersic function, the corresponding index will be denoted $n_b$.} \citep{dri06,bal06} and global color-concentration (log $n$) \citep{dri06}.  Importantly, many galaxies also display distinct, identifiable structural components---frequently (in massive systems) a central \textit{bulge} and a surrounding \textit{disc} \citep{dev59,fre70}, as well as bars, nuclei, and spiral arms.

Hierarchical clustering \citep{whi78,whi91,kau99,som99}, the predominant theory of galaxy formation, postulates the existence of two distinct structural formation mechanisms---the cooling of gas inside rotating dark matter halos to form \textit{discs} \citep{fal80} and the merging of similar-sized discs to form \textit{spheroids} (i.e., ellipticals and classical \textit{bulges}) \citep{too72,bau96}.  Hence, understanding the role of structure in shaping the observed global bimodality may prove crucial to understanding a range of galaxy formation processes.

Using the Millennium Galaxy Catalogue (MGC, \citealt{lis03}) of 10,095 galaxies, \citet{dri06} revealed a clear relationship between visual morphological type and position in the global color-concentration plane.  Specifically, E/S0s dominate the red, high-$n$ (centrally-concentrated) peak, Sd/Irrs dominate the blue, low-$n$ (diffuse) peak, and Sa-Sc galaxy types span both.  The authors proposed that this relationship derives from the two-component nature of galaxies, and that objects falling between the two peaks of the global bimodality result from the mixing of separate red, compact, bulge and blue, diffuse, disc components in varying degrees.  \citet{dro07} challenged this hypothesis based upon a sample of 39 S0-Sc galaxies with high-resolution, HST imaging.  Specifically, they recovered a correlation between galaxy color and (visually-classified) bulge type, such that classical bulge systems are globally red (i.e., also have red discs) and pseudobulge systems are globally blue, regardless of the bulge-to-total flux ratio.

Here we further explore the origin of the color-concentration bimodality using the results of Allen et al.'s (2006) bulge-disc decomposition of 10,095 bright galaxies in the MGC.  In Section \ref{data} we review the MGC structural dataset and describe the construction of a robust sample, supplemented with rest-frame component colors.  In Section \ref{results} we contrast the global color-concentration distributions of one-component and two-component galaxies; identify key structural types and construct their luminosity functions; and examine the relationship between bulge and disc colors.  Finally, in Section \ref{discussion} we summarize our conclusions, discuss our results in light of contemporary galaxy formation theory, and reflect upon the challenges facing low resolution, automated structural decomposition studies.  Unless otherwise stated, all magnitudes are given in the AB system and a cosmological model with $\Omega_M = 0.3$, $\Omega_\Lambda = 0.7$ and $H_0 = 70$ km s$^{-1}$ Mpc$^{-1}$ is used throughout.\\

\section{Data: The Millennium Galaxy Catalogue}\label{data}
The Millennium Galaxy Catalogue (MGC) is a deep ($\mu_{\mathrm{lim},B} = 26$ mag arcsec$^{-2}$), wide-field (30.88 deg$^2$), $B$-band imaging survey obtained with the Wide Field Camera on the 2.5-m Isaac Newton Telescope \citep{lis03}.  It is fully contained within both the 2dFGRS \citep{col01} and SDSS (DR5, \citealt{ade07}), which provide photometric information in additional filters and $\sim$4,700 high quality, spectroscopic redshifts.  A follow up campaign of spectroscopic observations (described in \citealt{dri05}) has ensured over 96\% redshift completeness for a sub-sample (MGC-BRIGHT) of 10,095 galaxies with $B < 20$ mag.

\subsection{Bulge-Disc Decomposition}
\citet{all06} modeled the 2D light distributions of the entire MGC-BRIGHT sample, providing a publicly available structural catalogue\footnote{http://www.eso.org/$\sim$jliske/mgc/}.  The 2D light distribution of each galaxy was fit with both a global S\'ersic (one-component) model and a S\'ersic bulge plus exponential disc (two-component) model\footnote{Specifically, the S\'ersic function describes a major axis intensity profile of the form: $I_b (R) = I_e \exp \left\{ -b_n \left[ (R/R_e)^{1/n}-1\right]\right\}$ where the effective radius, $R_e$, encloses half the component luminosity, and $I_e$ is the intensity at the effective radius.  The exponential disc model may be considered a specific case of the S\'ersic profile for $n=1$ with the following major axis intensity profile: $I_d (R) = I_0 \exp (-R/h)$ where $I_0$ is the central intensity and $h = 1.678 R_e$ is the disc scale-length.  For a comprehensive review of the S\'ersic profile see \citet{gra05}.} using the GIM2D analysis package \citep{sim98}.  In addition, \citet{all06} designed a logical filter to identify and replace illogical fits (e.g.\ inverted profiles where the inner bulge was fit by the exponential component and the outer disc by the S\'ersic component).  Based on the best-fit model parameters, the logical filter also offers an objective and quantitative subdivision of the local galaxy population into three morphological classes.  Namely, `bulge-plus-disc', for which the two-component model is preferred; and `elliptical' and `disc-only', for which the one-component model is preferred.  The distinction between the latter two classes is simply made according to the value of the best-fit global S\'ersic index (i.e., greater than or less than 1.5 respectively), recalling the absence of dwarf galaxies in MGC-BRIGHT.

\subsubsection{Treatment of Elliptical Galaxies}\label{false}
A known problem exists for automated photometric decomposition studies concerning the treatment of elliptical galaxies.  Namely, that two-component fits to eyeball-morphological ellipticals will readily converge upon a solution containing an outer disc, although no such structural component is actually present (see \citealt{tru04,gut04,all06}).  Rather, these `false discs' serve only to correct slight deviations in the measured light profile from a pure global S\'ersic model.  The origin of these deviations may be real (e.g.\ a recent merger event, twisting of the inner isophotes, or the presence of an outer halo) or artificial (e.g.\ incorrect estimation of the sky background).

\citet{all06} attempt to remove `false disc' systems from their two-component population via the logical filter, which assigns all galaxies with model fits corresponding to bulge-to-total flux ratios (B/T) greater than 0.8 to the one-component population.  The authors claim to demonstrate the success of this approach by contrasting the inclination distribution of their real disc sample, which is roughly consistent with a random distribution of projections on the sky, against that of their false disc sample, which is biased towards a more face-on distribution.

However, the B/T $>$ 0.8 cut employed by \citet{all06} is significantly less restrictive than the B/T $>$ 0.6 cut more commonly employed in contemporary studies (e.g.\ \citealt{tru04,gut04}).  Hence, we conducted visual inspection of the real (MGC) and model (GIM2D), $B$-band images and surface brightness profiles of 100 bright ($B < 19$ mag) galaxies from Allen et al.'s (2006) bulge-plus-disc sample in order to gauge the reliability of this dataset.  A total of 23 probable `false discs' were thereby identified amongst 72 real discs and 5 systems of indeterminate type.  Three examples of these `false disc' candidates are displayed in Fig.\ \ref{cut} in the Appendix.  These galaxies display elliptical visual morphologies, although slight flux asymmetries are revealed upon inspection of the image residuals from their one-component fits.  Such asymmetries may derive from a diverse range of factors, including recent mergers and the presence of undigested satellites.  

\begin{figure}
\epsscale{1.1}
\plotone{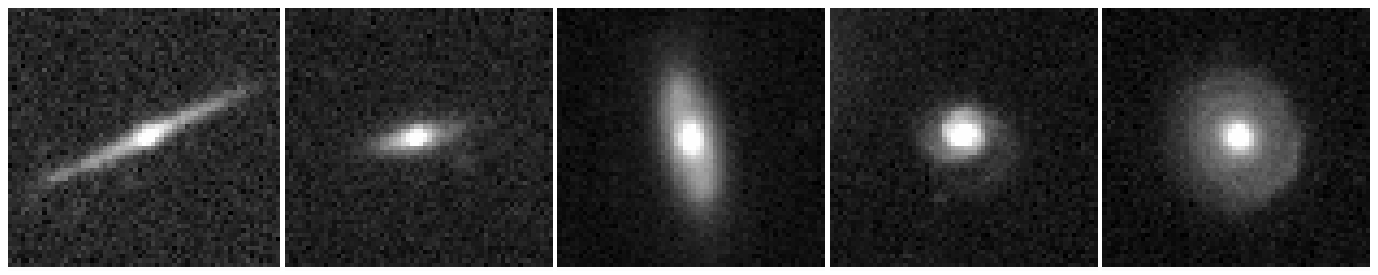}
\epsscale{1.1}
\plotone{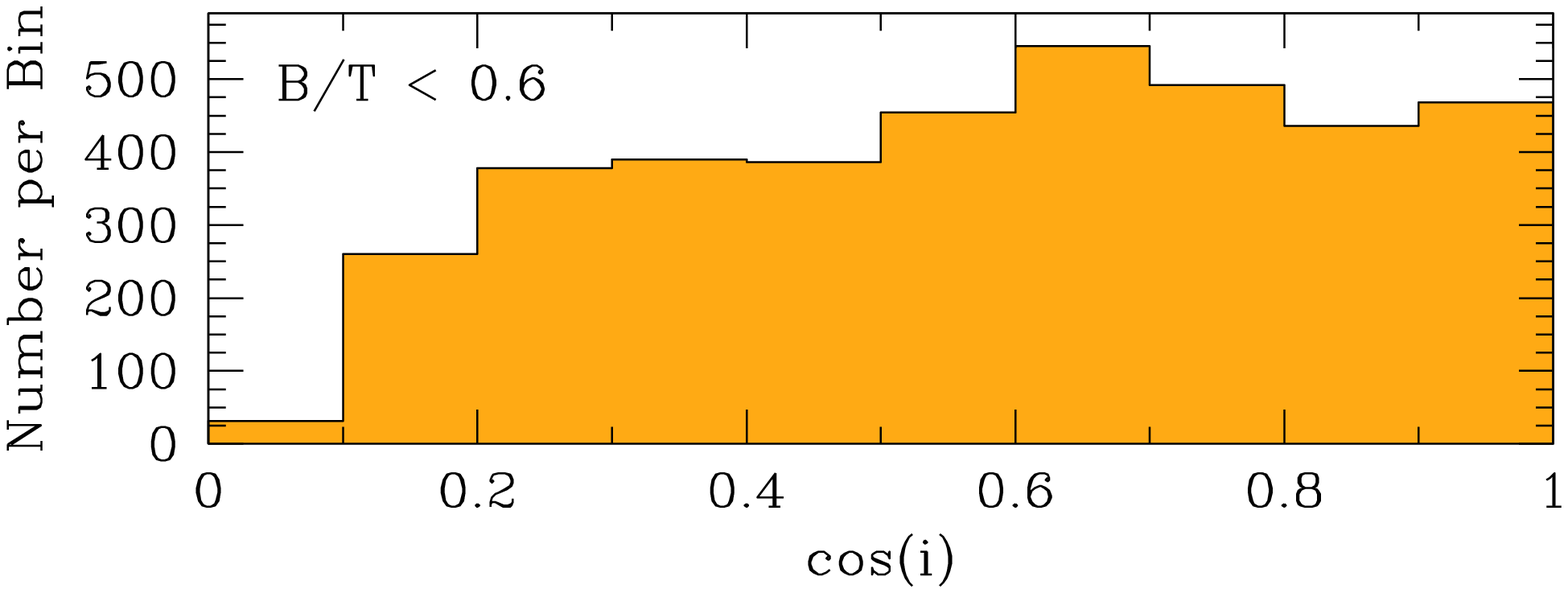}
\epsscale{1.1}
\plotone{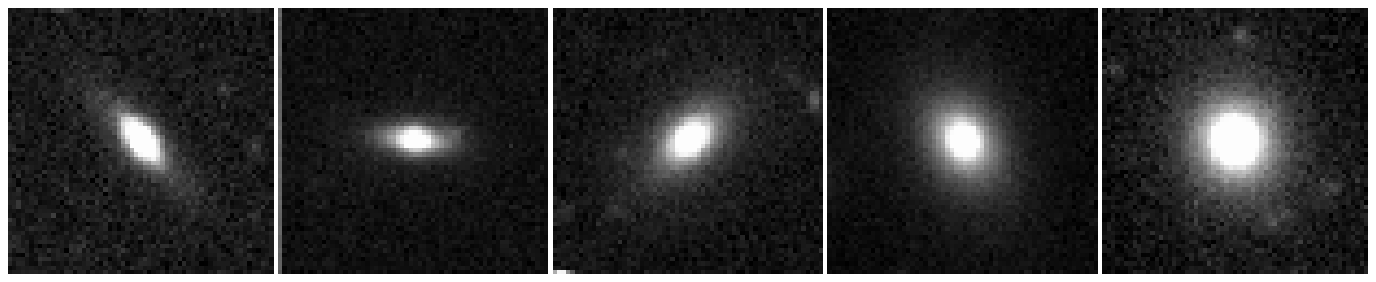}
\epsscale{1.1}
\plotone{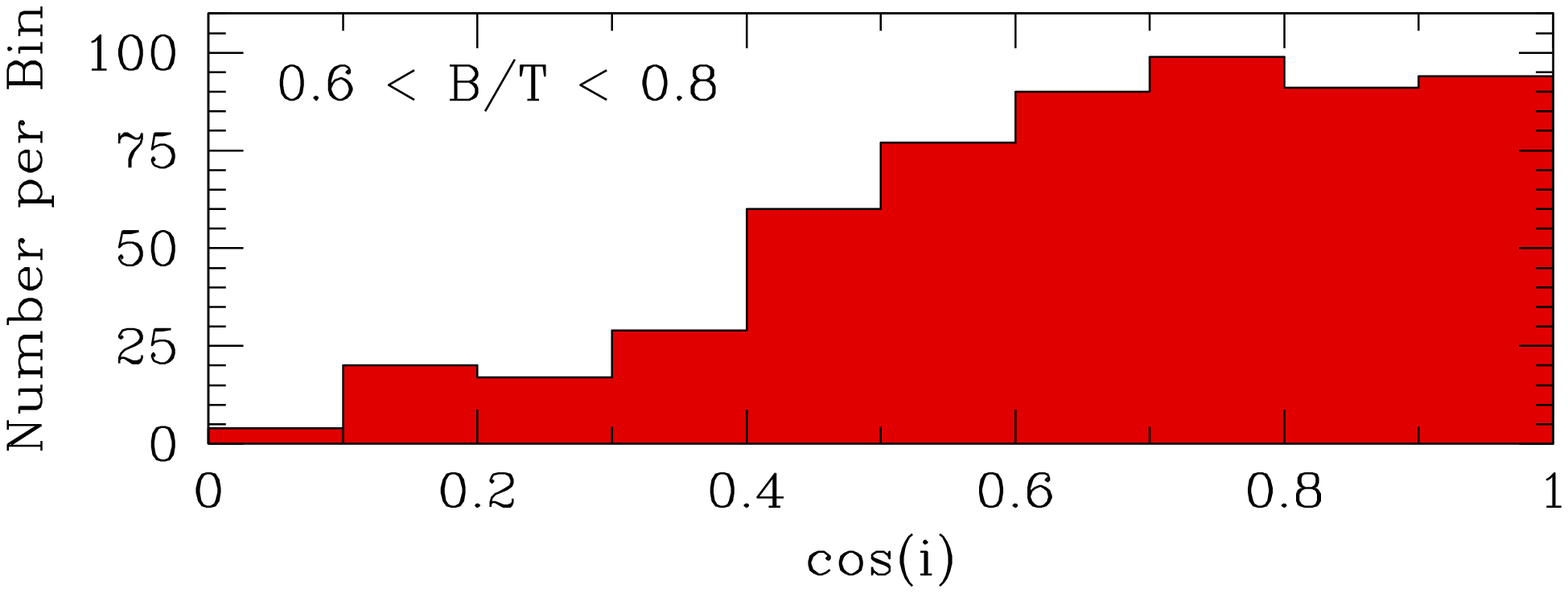}
\caption{\label{bt}The inclination distribution of galaxy `discs' in Allen et al.'s (2006) bulge-plus-disc class for systems with B/T $<$ 0.6 (\textit{top panel}) and 0.6 $<$ B/T $<$ 0.8 (\textit{bottom panel}).  The former is consistent with a random distribution of projections on the sky, while the latter is biased towards a more `face-on' distribution, indicative of `false disc' fits.  Postage-stamp, MGC $B$-band images of example galaxies from each B/T interval---spanning a range of inclinations---are also presented for comparison.  The lower B/T flux ratio systems display lenticular and spiral morphologies, whereas the higher B/T flux ratio systems generally display elliptical morphologies.}
\end{figure}

Reviewing the inclination distribution of Allen et al.'s (2006) bulge-plus-disc sample (their Fig.\ 9), we note that the logical filter has reduced, \textit{but not entirely removed}, the bias towards face-on `disc' fits.  Interestingly, this bias can be further reduced by applying the more conventional B/T $> 0.6$ cut (as shown in our Fig.\ \ref{bt}).  Of the 100 visually-classified systems discussed above, 15 of the 23 `false disc' candidates and only 2 of the 72 real discs are removed by this cut---revealing this to be a robust, yet conservative, step towards reducing `false disc' contamination.  All galaxies from Allen et al.'s (2006) bulge-plus-disc sample with B/T $>$ 0.6 (581 of 4427 objects) were therefore reassigned to the one component population for the remainder of this investigation.

\subsection{Sample Selection}
Motivated by Driver et al.'s (2005) study of the MGC luminosity-size plane, we select only those galaxies with \textsc{(i)} $0.013 < z < 0.18$ to ensure minimal impact from peculiar velocities, galaxy evolution, and the K-correction, and \textsc{(ii)} $\left<\mu\right>_{e,B} < 25.25$ mag arcsec$^{-2}$ to exclude the lowest signal-to-noise systems for which the photometry is less reliable.  We also restrict our analysis to near face-on discs ($i < 60$ deg) to minimise potential biases due to internal, inclination dependent dust extinction \citep{tuf04,dri07b}.  A limit of $M_B - 5 \log h_{70} < -18$ mag is then used to isolate the well-sampled bright-end of the luminosity function.

By duplicating their bulge-disc decomposition analysis on a sub-sample of 682 galaxies with repeat observations, \citet{all06} establish limits on the robustness of their derived \textit{global} structural parameters against variations in the local observing conditions.  Namely, for the accurate recovery of \textit{global} properties $R_e > 0.5\Gamma$ is required (where $R_e$ is the global half light radius and $\Gamma$ the characteristic seeing FWHM); and for the accurate recovery of \textit{bulge} properties $R_{e,b} > 0.5\Gamma$ is also required (where $R_{e,b}$ is the bulge half light radius).  Hence, we apply the former limit to our entire sample used to construct global color-concentration distributions, and we employ the latter limit to identify bulge-plus-disc systems with `confident' bulge S\'ersic indices (and hence B/T flux ratios) during the computation of component colors (see Section \ref{compcolors}).

In our final sample there are 917 elliptical, 1544 disc-only, and 1526 bulge-plus-disc systems (812 with `confident' bulge S\'ersic indices).  Visibility-volume weights \citep{phi90} were computed to account for the relevant observational biases against each structural type.  These weights are used throughout this paper to recover volume-density distributions from the raw data.

\subsection{Global and Component Colors}\label{compcolors}
All galaxies in MGC-BRIGHT have well-matched counterparts in the SDSS-DR5 \citep{ade07}, which provides PSF, Petrosian, and model magnitudes in the Sloan $u$$g$$r$$i$$z$ filter system \citep{gun98}.  We compute (apparent) \textit{global} $(u-r)_g$ colors for our entire sample using the SDSS model magnitudes because these offer the highest signal-to-noise measurements, particularly for the problematic $u$-band (see \citealt{bal04}).

We also derive (apparent) \textit{component} $(u-r$) colors for our bulge-plus-disc systems as follows.  Core colors are computed from the SDSS PSF magnitudes via the relation $(u-r)_{\mathrm{core}} = u_\mathrm{PSF} - r_{\mathrm{PSF}} + 5 \log_{10}(\Gamma_u/\Gamma_r)$.  The third term in this equation, $5 \log_{10}(\Gamma_u/\Gamma_r)$, is required in order to account for an intrinsic difference in area (and hence, enclosed core flux) between the SDSS $u$ and $r$-band PSFs.  We then equate this core color with the color of the bulge, which is a reasonable assumption provided the bulge is a similar size to the SDSS $u$ and $r$-band PSFs ($\Gamma_{u} \approx 1.6$ and $\Gamma_{r} \approx 1.4$ arcsec respectively).  For bulges much smaller than the SDSS PSFs the core color may be biased by the inclusion of significant disc flux.  For bulges much larger than the SDSS PSFs use of the core color may give undue weight to small-scale, radial variations in the bulge color resulting from compact central nuclei, star-forming rings, nuclear bars, and/or nuclear dust lanes \citep{jog99,car01,fis06}.  The radius inside which the model bulge intensity exceeds that of the model disc (i.e., the `anchor point', $r_\mathrm{anc}$) provides a useful measure of bulge size in this context.   Of our 812 component bulges with `confident' S\'ersic indices, 145 are a similar size to the SDSS $u$ and $r$-band PSFs (i.e, $0.75 < r_{\mathrm{anc}}/\left<\Gamma_u,\Gamma_r\right> < 1.5$), while 217 are smaller and 450 are larger.  Hence, a distinction is made in our later analysis between galaxies with `confident' and `less confident' bulge (and disc) colors based on this size criterion.

We then suppose that the (unknown) $g$-band B/T flux ratios of our two-component systems may be approximated by their (known) $B$-band values, given the overlapping wavelength ranges covered by these two filters.  This enables the computation of $g$-band bulge magnitudes from the SDSS $g$-band model magnitudes, and thus $u$ and $r$-band bulge magnitudes using our bulge colors.  The corresponding disc magnitudes in these Sloan filters are then obtained  by subtracting the bulge flux from total (model) flux, and disc colors derived from their difference.  The relevant formula is 
\begin{eqnarray} &&(u-r)_d =     \\  &&-2.5\log_{10} \left( \frac{ 10^{-0.4 u_\mathrm{model}} - \frac{B}{T} 10^{-0.4(g_\mathrm{model} - (g-u)_{b}) }}{10^{-0.4 r_\mathrm{model}} - \frac{B}{T} 10^{-0.4(g_\mathrm{model} - (g-r)_{b} )}} \right). \nonumber
\end{eqnarray}

K- and e- corrections are required for the computation of \textit{absolute} magnitudes and rest-frame colors. Individual K-corrections for galaxies in MGC-BRIGHT were computed by \citet{dri05} by matching broadband SEDs provided by the (global) MGC and SDSS magnitudes to the 27 spectral templates given in \citet{pog97}.  Here we repeat this process in order to generate accurate and consistent K-corrections for the bulges and discs of our two-component systems using the separate colors derived above.  For the e-correction \citet{dri05} adopt a pure luminosity evolution model of the form \begin{equation} L_{z=0} = L_z (1+z)^{-\beta} \end{equation} where $\beta=0.75$ in the $B$-band, regardless of galaxy type.  For consistency with earlier work we continue to utilize the MGC $B$-band absolute magnitudes derived in this manner, but extend this approach to better estimate rest-frame ($u-r$) colors.  Specificially, we adjust the $\beta$ values for each galaxy and component according to its best-fit spectral type.  \citet{pog97} employ three stellar population models, roughly corresponding to E, Sa and Sc spectral types, as the basis of their 27 templates.   The low redshift evolution in these models may be appoximated by $\beta$ values of 1.1, 1.4, and 2.2 in the $u$-band and 0.82, 0.92, and 1.1 in the $r$-band respectively.

\subsection{Dust}\label{dust}
Accounting for the impact of both internal and external dust extinction is an essential step towards the recovery of colors and luminosities representative of galaxy stellar populations.  All MGC magnitudes \citep{lis03} are corrected for \textit{external} dust extinction (i.e., that arising due to dust within the Milky Way) using the maps of \citet{sch98}.  

Recently, \citet{dri07b} recovered empirical inclination-attentuation relations from the MGC structural catalogue, and used these to calibrate opacities in Tuffs et al.'s (2004) multi-component dust models.  The mean dust corrections thereby derived (see Table 1 of \citealt{dri08}) are employed in this study to account for the effects of \textit{internal} dust extinction on various disc galaxy observables.   For our inclination-limited sample ($i < 60$ deg) we obtain median corrections of $\Delta (u-r)_d = -0.28$ mag and $\Delta M_B = -0.27$ mag for component discs, and $\Delta (u-r)_b = -0.20$ mag and $\Delta M_B = -0.98$ for component bulges.   These corrections are applied to all our disc-only and bulge-plus-disc systems, and their impact is indicated via reddening vectors marked on the relevant plots in Section \ref{results}.

However, observations indicating a possible variation in dust content with morphological type raise questions regarding the validity of these corrections for the entire bulge-plus-disc galaxy population.  In particular, a strong decrease is observed in the visually-estimated dust content of galaxies from late to early Hubble types \citep{van07}, although far-IR emission estimates of dust content suggest a weaker trend \citep{vla05}.  As Tuffs et al.'s (2004) dust model was devised with reference to nearby spiral galaxies only, and Driver et al.'s (2007b) inclination attenuation relation is dominated by late-type discs, it may not be appropriate to apply the resulting corrections to early-type discs.  Although, some recent observations suggest a surprising level of similarity between the \textit{distributions} of dust in these morphological types, including the presence of `dust spiral arms' in S0s \citep{pah04}.  In Section \ref{results} we discuss this potential source of bias in our results, and compute alternative luminosity functions for early-type bulge-plus-disc systems.\\

\subsection{Measurement Uncertainties}
Allen et al.'s (2006) catalogue includes 68\% confidence limits on the measurement of each structural parameter, estimated by GIM2D based on the shape of the $\chi^2$ minimum in which each fit converges.  The median error on the global S\'ersic indices of galaxies in our sample is $\Delta \log n = \pm 0.12$.  Measurement uncertainties on our absolute $(u-r)$ colors may be estimated as follows.  Firstly, we note that for all galaxies in our sample the relevant SDSS magnitudes have quoted errors of less than $\pm0.01$ mag.  Secondly, we observe that the vast majority of our global and component, broadband SEDs are well-matched by one of the 27 spectral templates from \citet{pog97}, and thus adopt a maximum uncertainty of $\pm$1 Hubble morphological type.  This translates to K-correction and e-correction error contributions to the absolute $(u-r)$ colors of no more than $\pm$0.03 mag and $\pm$0.14 mag respectively.  A further error of $\pm0.09$ mag is contributed by uncertaintes in the internal dust corrections derived by \citep{dri07b} and the measurement of disc inclinations.  Hence, the total uncertainty on our absolute $(u-r)$ colors is roughly $\pm 0.17$ mag.  Of course, the uncertainties in our `less confident' \textit{component} bulge and disc colors will be significantly larger, so we remove these systems from our analysis of component colors in Section \ref{bdcolors}.

\subsection{The Impact of Unmodeled Structural Features}\label{barbiases}
The presence of unmodeled structural features in a number of our two-component systems contributes an additional source of uncertainty in the measurement of bulge and disc colors which is not accounted for by our limits on size used to identify galaxies with `confident' bulge S\'ersic indices and colors.  Specifically, a number of authors (e.g.\ \citealt{lau07,gad08a,wei08}) have recently demonstrated that neglecting to account for bars during bulge-disc decompositions can lead to large systematic biases: bulge S\'ersic indices, scale-sizes, and fluxes are over-estimated as the model bulge component is forced wider to account for bar flux, and the corresponding disc profile becomes steeper to compensate \citep{gad08a}.  From a sample of 186 bright nearby galaxies, \citet{esk00} estimate (via visual classification) that over half of all bright spirals are strongly-barred in the near-IR, although this fraction drops to roughly a third in the optical.  Hence, bars are clearly an important feature to consider during structural decompositions.

As a demonstration of the impact of bars on Allen et al.'s (2006) bulge-disc decomposition results we derive alternative fits to a number of (visually-classified) strongly-barred galaxies in our two-component sample via multi-component (S\'ersic bulge plus S\'ersic bar plus exponential disc) fits with GALFIT version 2.0.3c \citep{pen02}.  We note that while the truncated S\'ersic profiles offered by the rival BUDDA software package \citep{des04} potentially enable a more faithful representation of bar light profiles (and often provided the most satisfactory fits to MGC galaxies in informal tests), only GALFIT currently allows the input of user-generated PSF functions necessary for fair comparisons with Allen et al.'s (2006) GIM2D analysis.  The real (MGC) and model (GIM2D and GALFIT), $B$-band images and surface brightness profiles of three example galaxies in which the addition of a bar component offered a superior fit are shown in Fig.\ \ref{bars} in the Appendix.  In each case the multi-component model fit returned a much lower bulge S\'ersic index and B/T flux ratio than in the original, two-component model fit.  Thus, we recommend future, automated, bulge-disc decomposition studies incorporate the modeling of bars from the onset, despite the added complexity thereby introduced into quality control and interpretation of the output.  We intend to explore the nature of bars in the MGC further in an upcoming paper (Cameron et al., in prep.).

Finally, we note that late-type discs with bright knots of active star-formation also create a class of problematic fits.  Examples of the real (MGC) and model (GIM2D and GALFIT), $B$-band images and surface brightness profiles of three such galaxies are displayed in Fig.\ \ref{starform} in the Appendix.  A comparison is again made to alternative fits using GALFIT, but this time with only a two-component (S\'ersic bulge plus exponential disc) model as these objects appear unbarred.  Despite the use of identical galaxy and PSF models, GALFIT and GIM2D return significantly different bulge structural parameters in two of these examples---a reflection of their differing fit algorithms.  The presence of complex, irregular structure precludes clean fits with smooth elliptical isophote models, and identification of the global minimum becomes difficult.  In the examples shown, neither software package offers satisfactory output as each has accounted for some intermediate radius star-formation flux via the bulge model.  The up-coming version of GALFIT (v3.0) will introduce features for mapping asymmetric, complex structure within the context of well-defined radial surface brightness profiles, and thus offers an exciting possibility for improved 2D bulge-disc decomposition of such galaxies.

Importantly, however, extensive visual inspection of galaxies in each of our key structural classes (i.e., elliptical, bulge-plus-disc, and disc-only) indicates that the presence of unmodeled structural components very rarely results in a clear mis-classification.  Hence, we conclude that, although \textit{component} bulge S\'ersic indices and B/T flux ratios (and hence colors) for some fraction of our galaxies may be systematically biased, the presence of unmodeled bars and bright knots of active star-formation in the MGC $B$-band images will \textit{not} affect our investigation of the role of galaxy structural type in shaping the \textit{global} color-concentration bimodality.

\section{Results}\label{results}
\subsection{The Global Color-Concentration Distributions of One- and Two-Component Galaxies}\label{results_global}
\begin{figure}
\center
\epsscale{1.1}
\plotone{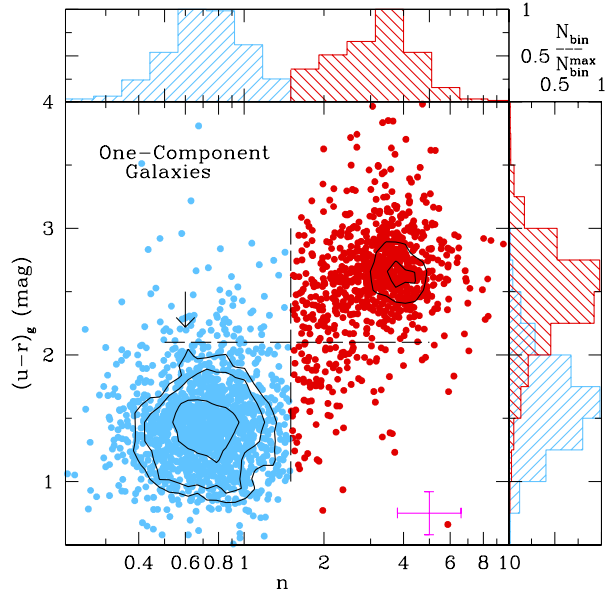}
\caption{\label{global}The global color-S\'ersic index distribution of bright ($M_B - 5 \log h_{70} < -18$ mag), nearby ($0.013 < z < 0.18$), one-component galaxies in the MGC.  Disc-only systems are represented by \textit{filled, blue circles} and elliptical systems by \textit{filled, red circles} with the subdivision at $n=1.5$ used to define these structural types indicated by the \textit{vertical, long-dashed, black line}.  The corresponding subdivision in color at $(u-r)_g=2.1$ mag is indicated by the \textit{horizontal, long-dashed, black line}.  The volume-density distribution of the one-component population is revealed by the \textit{solid, black contours} marked at log $\phi$ $=-5.25$, $-5$ and $-4.75$ $h_{70}^3$ Mpc$^{-3}$ per bin with the parameter space shown divided into a 48$\times$48 square grid.   The median reddening correction for internal dust extinction applied to the disc-only systems is indicated with a \textit{black arrow}. The mean uncertainties in the measurements of global colors and S\'ersic indices for galaxies in our sample are indicated by the \textit{magenta} error bar at the bottom of this plot.  Normalized histograms of the distributions of elliptical and disc-only galaxies in each variable are contained in the \textit{sidepanels}.}
\end{figure}

The global color-concentration distribution of bright ($M_B - 5 \log h_{70} < -18$ mag), nearby ($0.013 < z < 0.18$), \textit{one-component} galaxies in the MGC is presented in Fig.\ \ref{global}.  This structural class displays strong bimodality, separating naturally in this parameter space into two distinct sub-populations---a `red, highly-concentrated' peak and a `blue, diffuse' peak---composed of elliptical and disc-only systems respectively.  These peaks are evidently the foundations of those observed in the global color-concentration distribution of the full MGC population (Fig.\ 15 in \citealt{dri06}), although their separation is more clearly defined in the absence of the two-component systems.  We quantify this improvement in the apparent bimodality of the distribution by comparing the reduced-$\chi^2$ values resulting from least-squares fitting of pairs of bivariate Gaussian functions to the observed global color-concentration distributions of \textit{one- and two-component galaxies} ($\chi^2 / \nu = 14.2$) and \textit{one-component galaxies only} ($\chi^2 / \nu = 2.1$)\footnote{Of course, neither of these reduced-$\chi^2$ values represents a technically good fit as expected given that both peaks are slightly assymetric in shape, indicating intrinsic deviations from simple bivariate Gaussian distributions.  We also note that a 2D K-S test comparing the one and two component galaxy color-concentration distributions returned a probabilty of $\ll$1\% that these samples were drawn from the same population.}.  The centroids and widths of each peak, as derived by this bivariate Gaussian fitting approach, are compiled in Table \ref{pars} for reference.

Inspection of Fig.\ \ref{global} confirms that the subdivision of the one-component population into elliptical and disc-only systems at $n=1.5$ adopted by \citet{all06} is indeed consistent with a local minimum in this parameter.  An equivalent local minimum in color exists at roughly $(u-r)_g = 2.1$ mag.  Relative to these subdivisions (indicated in Fig.\ \ref{global}), luminous, `bulge-less, red discs' and `disc-less, blue bulges' (i.e., blue spheroids) are rare; specifically, we recover total volume-densities of only $1.7 \pm 0.3$ and $1.1 \pm 0.1$ $\times$ $10^{-4}$ $h^{3}_{70}$ Mpc$^{-3}$ respectively in the MGC.  Implications of the paucity of these systems for galaxy formation scenarios are discussed in Section \ref{discussion}.

\begin{deluxetable*}{lccccc}
\tablecolumns{6}
\tablecaption{Centroids and Standard Deviations of the One-Component Galaxy Color-Concentration Distribution Peaks\label{pars}}
\tablehead{\colhead{} & \colhead{$ \left< n \right>$} & \colhead{$\sigma_{\log n}$} & \colhead{$ \left< (u-r)_g \right> $} & \colhead{$\sigma_{(u-r)_g}$}\\
\colhead{} & \colhead{} & \colhead{} & \colhead{(mag)} & \colhead{(mag)}}
\startdata
Elliptical & $3.4$\phantom{1} $\pm$ $0.2$\phantom{0} & $0.15 \pm 0.01$ & $2.62 \pm 0.09$ & $0.27 \pm 0.01$\\
Disc-only & $0.71$ $\pm$ $0.02$ & $0.16 \pm 0.01$ & $1.44 \pm 0.04$ & $0.28 \pm 0.04$\\
\enddata
\tablecomments{Values quoted are relative to the absolute magnitude limit of the sample ($M_B - 5 \log h_{70} < -18$ mag), recalling that mean red and blue sequence colors \citep{bal04}, and elliptical galaxy S\'ersic indices \citep{gra03}, are luminosity dependent.}
\end{deluxetable*}

\begin{figure}
\center
\epsscale{1.1}
\plotone{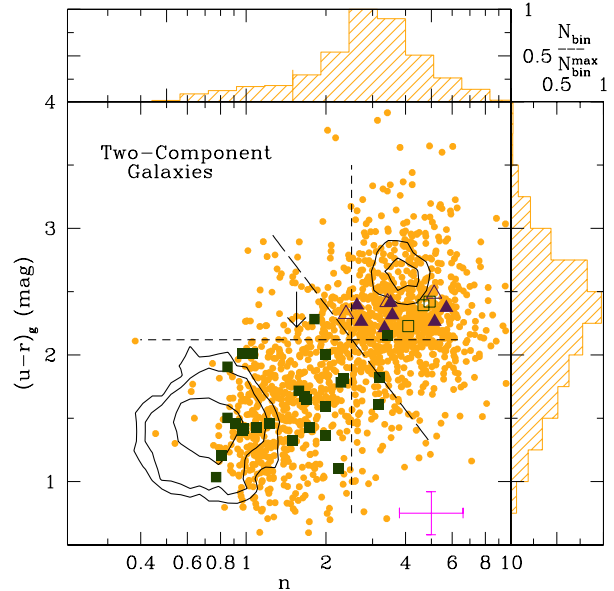}
\caption{The global color-S\'ersic index distribution of bright ($M_B - 5 \log h_{70} < -18$ mag), nearby ($0.013 < z < 0.18$), two-component galaxies in the MGC.  These bulge-plus-disc systems are represented by \textit{filled, orange circles}.  For reference, the volume-density distribution of the one-component population is revealed by the \textit{solid, black contours} marked at log $\phi$ $=-5.25$, $-5$ and $-4.75$ $h_{70}^3$ Mpc$^{-3}$ per bin with the parameter space shown divided into a 48$\times$48 square grid.   The median reddening correction for internal dust extinction applied to these bulge-plus-disc systems is indicated with a \textit{black arrow}.  Drory \& Fisher's (2007) sample of 39 S0-Sc galaxies (with colors adjusted for dust reddening and a magnitude system offset, as described in Section \ref{results_global}) is overlaid for comparison.  Their (visually-classified) pseudobulge galaxies are marked as \textit{green squares} and their classical bulge galaxies as \textit{purple triangles} with \textit{open} symbols denoting S0s and \textit{solid} symbols Sa-Sc types.  Our subdivision of the two-component population into `bridging' and `red peak' structural types at $(u-r)_g = 3.22-2.75\log(n)$ is indicated by the \textit{long-dashed, black line}.  For comparion, Drory \& Fisher's (2007) subdivisions at $n=2.5$ and $(u-r)_g = 2.2$ mag (adjusted to $2.1$ mag) are indicated by the \textit{short-dashed, black lines}.  The mean uncertainties in the measurements of global colors and S\'ersic indices for galaxies in our sample are indicated by the \textit{magenta} error bar at the bottom of this plot.  Normalized histograms of the two-component galaxy distribution in each variable are contained in the \textit{sidepanels}.\label{twocomp}}
\end{figure}

The global color-concentration distribution of bright ($M_B - 5 \log h_{70} < -18$ mag), nearby ($0.013 < z < 0.18$), \textit{two-component} galaxies is presented in Fig.\ \ref{twocomp}.  These bulge-plus-disc systems constitute an intermediate color-concentration class, spanning the two peaks of the one-component galaxy bimodality.   As mentioned earlier, \citet{dro07} identified a relationship between visually-classified bulge type (i.e., `pseudobulge' or `classical' bulge)\footnote{Although visual bulge morphology has been shown to correlate with $V-H$ color \citep{car02} and mid-infrared bulge color \citep{fis06} we would caution that dynamical data (indicating the degree to which these structures are rotationally-supported or otherwise) would allow for more confident classification of bulge type.} and position in the color-concentration plane.  Their sample of 39 S0-Sbc galaxies is overlaid against the MGC distribution in Fig.\ \ref{twocomp} for comparison, and a close agreement is evident.  (Note that we adjust the colors quoted in Drory \& Fisher's (2007) Table 1 for the effects of internal dust reddening, and for an offset of $\sim$0.2 mag between their integrated S\'ersic profile colors and the SDSS model colors used here.)  \citet{dro07} reveal that Sabc, pseudobulge galaxies may be isolated from S0/Sabc, classical bulge (and S0 pseudobulge) galaxies in this parameter space using cuts at $n=1.5$ and $(u-r)_g = 2.2$ mag ($\sim$2.1 mag after the relevant corrections).  Motivated by these results we subdivide our two-component population into two structural types, `bridging' and `red peak', defined via a cut at $(u-r)_g = 3.22-2.75\log n$ (indicated in Fig.\ \ref{twocomp}).  Visual inspection of the $B$-band MGC images confirms these structural types correspond to morphologically late-type and early-type disc respectively, as reported by \citet{dro07}.

Examining the two-component, red peak systems we note that their global colors are significantly bluer than those of the one-component, red peak (i.e., elliptical) systems, which appears to indicate a younger mean stellar population age for these galaxies (modulo the effects of metallicity).  However, this offset is almost entirely introduced by our correction for dust reddening, which (as noted earlier) may not be appropriate for these early-type discs.  Hence, we caution against placing excessive emphasis on this color offset in the interpretation of our results.  We also note that the two-component, red peak systems display highly-concentrated light distributions similar to those of the ellipticals.  Although this appears to argue for the role of violent-relaxation processes during their formation, the known prevalence of (secular evolution-built) pseudobulges amongst early-type disc galaxies \citep{erw03,lau05,dro07} challenges this interpretation.

The two-component, bridging population systems are generally similar in color to their one-component, disc-only couterparts, albeit slightly redder, whilst their global S\'ersic indices are significantly higher.  These results are broadly consistent with the interpretation of these galaxies as pseudobulge systems given that N-body simulations demonstrate the effectiveness of secular evolution in building up comparable central mass concentrations over 1-2 Gyr timescales (e.g.\ \citealt{ath05,deb06}).  We will return to the interpretation of our results in light of galaxy formation theory in Section \ref{discussion}---but first we recover luminosity functions and total stellar mass densities for each structural type (Section \ref{results_lfs}), and explore the connection between bulge and disc colors in two-component systems (Section \ref{bdcolors}).

\begin{deluxetable*}{lccccc}
\tablecolumns{9}
\tablecaption{Luminosity Function Parameters for Key Structural Types in the MGC\label{lfpars}}
\tablehead{\colhead{} & \colhead{$M^{\ast}_{B}-5\log h_{70}$} & \colhead{$\alpha$} & \colhead{$\phi_{\ast}$} & \colhead{$j_B$} & \colhead{$\rho$}\\
\colhead{} & \colhead{(mag)} & \colhead{} & \colhead{($10^{-3}$ $h^{3}_{70}$ Mpc$^{-3}$ $[0.5$ mag$]^{-1}$)} & \colhead{($10^{8}$ $h_{70}$ $L_{\odot}$ Mpc$^{-3}$)} & \colhead{($10^{8}$ $h_{70}$ $M_{\odot}$ Mpc$^{-3}$)}}
\startdata
One-Component (All) & $-19.05 \pm 0.07$ & $-1.03 \pm 0.04$ & $5.8 \pm 0.2$ & \phn$1.3 \pm 0.1$\phn & $2.0 \pm 0.2$\\
\phn \phn Disc-only\dotfill & $-19.14 \pm 0.07$ & $-1.08 \pm 0.04$ & $4.0 \pm 0.2$ & \phn$1.0 \pm 0.1$\phn & $1.3 \pm 0.1$\\
\phn \phn Elliptical\dotfill  & $-18.93 \pm 0.08$ & $-0.69 \pm 0.04$ & $1.5 \pm 0.1$ & $0.27 \pm 0.07$ & $0.7 \pm 0.1$\\
Two-Component (All) & $-19.82 \pm 0.08$ & $-0.61 \pm 0.04$ & $3.3 \pm 0.2$ & $1.3 \pm 0.1$\ & $2.9 \pm 0.3$\\
\phn \phn Red peak\dotfill  & $-19.69 \pm 0.09$ & $-0.20 \pm 0.05$ & $1.6 \pm 0.1$ & $0.59 \pm 0.06$ & $1.8 \pm 0.2$\\
\phn \phn Red peak (non-DC)\dotfill  & $-19.10 \pm 0.08$ & $-0.39 \pm 0.05$ & $1.8 \pm 0.1$ & $0.37 \pm 0.04$ & $1.4 \pm 0.2$\\
\phn \phn Bridging\dotfill  & $-19.59 \pm 0.08$ & $-0.84 \pm 0.04$ & $2.0 \pm 0.2$ & $0.68 \pm 0.07$ & $1.1 \pm 0.1$\\
\enddata
\tablecomments{All quoted uncertainties are the random errors in the Schechter function fits.  Based on mock 2dFGRS NGP catalogues \citep{col98}, we estimate that the potential systematic errors on these values due to cosmic variance amounts to 13\% \citep{dri05}.  A similar level of uncertainty exists in the elliptical, two-component (all) and red peak parameters due to the potential for mis-classification of ellipticals as bulge-plus-disc systems (see the discussion regarding `false discs' in Section \ref{false} and Fig.\ \ref{cut}).}
\end{deluxetable*}

\begin{figure}
\center
\epsscale{1.1}
\plotone{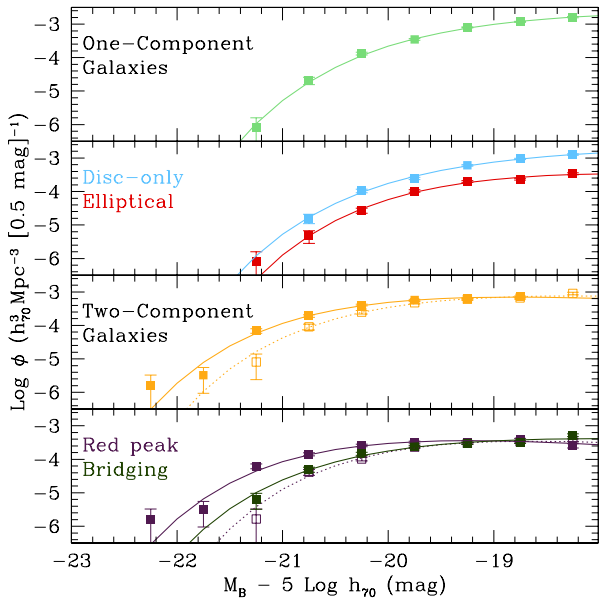}
\caption{\label{lfs}The $B$-band luminosity functions of key structural types identified in the MGC.  Namely, one-component (\textit{light green}) and two-component (\textit{orange}) systems, and their corresponding sub-types: elliptical (\textit{red}) and disc-only (\textit{blue}); bridging (\textit{dark green}) and red peak (\textit{purple}) (see Section \ref{results_global} for definition).  In each case, the measured number density of galaxies in each 0.5 mag bin is indicated by a \textit{solid square} with Poisson error bars, and the corresponding best-fit Schechter luminosity function by a \textit{solid line}.  The \textit{open squares} with error bars and \textit{dotted lines} illustrate the effect of removing the internal dust extinction correction applied to the two-component, red peak systems.  The Schechter function parameters of each fit are contained in Table \ref{lfpars}.}
\end{figure}

\subsection{Luminosity Functions and Stellar Mass Densities of Key Structural Types}\label{results_lfs}
The luminosity functions of the key MGC structural types identified in Section \ref{results_global}---namely, one-component (elliptical and disc-only) and two-component (bridging and red peak)---are presented in Fig.\ \ref{lfs}.  Each of these was modeled using a Schechter luminosity function with the best-fit Schechter parameters and integrated luminosity densities contained in Table \ref{lfpars}.

Inspection of the luminosity functions by structural type in Fig.\ \ref{lfs} reveals that two-component galaxies dominate (in number density) over one-component galaxies at the brightest luminosities ($M_B - 5 \log h_{70} < -20$ mag), whereas one-component galaxies dominate at intermediate luminosities ($-20 < M_B - 5 \log h_{70} < -18$ mag).  Furthermore, over the luminosity range explored, disc-only systems are more common than ellipticals, and red peak systems are more common than bridging systems.\footnote{However, if the dust attenuation corection is removed from the red peak systems (as discussed in Section \ref{dust}), bridging systems are more common at the brightest luminosities.}  No previous study has examined luminosity functions by structural type using an equivalent bulge-disc decomposition approach.  However, we note that our results are qualitatively consistent with previous luminosity functions by visual morphological class (given the expected morphological composition of each structural type).  In particular, the $B$-band luminosity functions compiled by \citet{del03} reveal that Sa/Sb (two-component, red peak and bridging) galaxies dominate at the brightest luminosities, while Sc (one-component, disc-only; and two-component, bridging) and Sd/Irr (one-component, disc-only) galaxies dominate at intermediate and faint luminosities respectively.

We also note that the total luminosity density of all structural types derived here, $j_B = 2.6\pm0.3 \times 10^{8}$ $h_{70}$ $L_\odot$ Mpc$^{-3}$ is similar to that for all structural components (i.e., bulges and discs) derived by \citet{dri07b}, $j_B = 3.0\pm0.2 \times 10^{8}$ $h_{70}$ $L_\odot$ Mpc$^{-3}$.  Both studies utilize the MGC structural catalogue, however, the total luminosity density obtained here is slightly smaller due largely to the reclassification of B/T $>$ 0.6 galaxies as ellipticals (intended to remove `false discs'; see Section \ref{false})---when returned to the elliptical population these galaxies are no longer dust corrected, and hence become less luminous.  We therefore also note the revised luminosity density of elliptical galaxies derived here, $j_B = 0.27\pm0.07 \times 10^{8}$ $h_{70}$ $L_\odot$ Mpc$^{-3}$, which is higher than that recovered by \citet{dri07b}, $j_B = 0.14\pm0.02 \times 10^{8}$ $h_{70}$ $L_\odot$ Mpc$^{-3}$.

We also compute stellar mass densities using the model relationships between $(u-r)$ color and mass-to-light ratios given by \citet{bel03}.  We thereby recover total stellar mass densities in each structural type of $0.7 \pm 0.1$, $1.3 \pm 0.1$, $1.8 \pm 0.2$ and $1.1 \pm 0.1$ $\times 10^{8}$ $h_{70}$ $M_\odot$ Mpc$^{-3}$ for elliptical, disc-only, red peak and bridging systems respectively.  Again we note that our total stellar mass density for all structural types, $4.9 \pm 0.3$ $\times 10^{8}$ $h_{70}$ $M_\odot$ Mpc$^{-3}$, is similar to that recovered by \citet{dri07b} for all structural components (i.e., bulges and discs) in the MGC, $5.2 \pm 0.4$ $\times 10^{8}$ $h_{70}$ $M_\odot$ Mpc$^{-3}$.

The comparable luminosity and mass densities of one- and two-component systems revealed here highlights the importance of considering the fundamental distinction between elliptical, disc-only and bulge-plus-disc systems in order to properly understand galaxy formation and evolutionary processes.  These luminosity functions by \textit{structural type}, and those by separate \textit{structural component} presented in \citet{dri07b}, offer valuable constraints for galaxy formation models, and should serve as key reference points for semi-analytical simulations.

\subsection{Relationship between Bulge and Disc Colors}\label{bdcolors}
\begin{figure}[!t]
\center
\epsscale{1.15}
\plotone{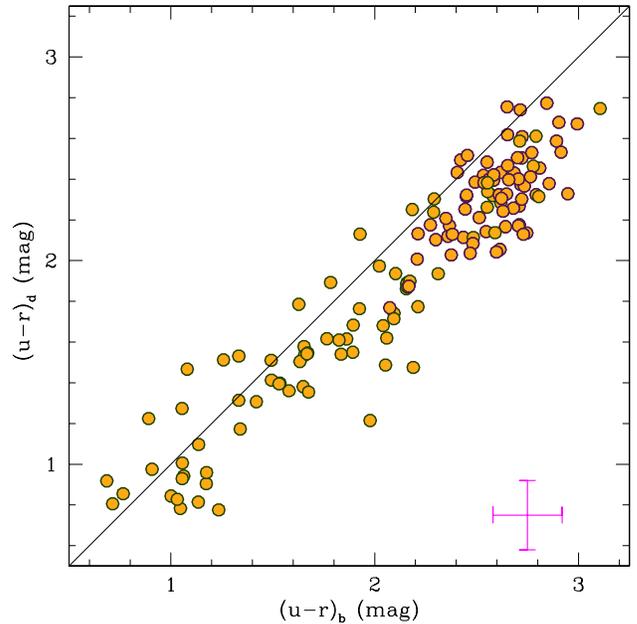}
\caption{\label{colors}The relationship between the bulge and disc colors of two-component galaxies in the MGC.  Only systems with `confident' \textit{component} colors are shown on this plot.  Each bulge is marked by a \textit{circle}, colored \textit{dark green} if the bulge is embedded in a bridging population galaxy and \textit{purple} if the bulge is embedded in a red peak population galaxy.  The estimated uncertainties in the measurements of bulge and disc colors for galaxies in our sample are indicated by the \textit{magenta} error bar.}
\end{figure}

In Section \ref{results_global} we demonstrated that the dichotomous disc-only and elliptical (or `bulge-only') galaxy populations are the foundations of the bulge, low-$n$ and red, high-$n$ peaks of the global color-concentration bimodality, and that bulge-plus-disc galaxies span these two peaks---as expected given the correlation between global color-concentration and eyeball morphological type reported by \citet{dri06}.  However, as noted by \citet{dro07}, Driver et al.'s (2006) hypothesis that the global properties of bulge-plus-disc systems are the result of mixing red, high-$n$ bulges and blue, low-$n$ discs in varying degrees is contradicted by the close relationship between bulge and disc colors \citep{pel96,mac04}.  Here we confirm this relationship for the bulge and disc colors of MGC two-component galaxies (with `confident' colors), as shown in Fig.\ \ref{colors}.  In Peletier \& Balcells' (1996) study of 30 highly-inclined, S0-Sbc galaxies the authors measured component colors via wedge-shaped apertures designed to avoid biases due to dust attentuation in the plane of each disc.  They thereby measured a (`dust-free') mean bulge-disc color offset of $\left< (U-R)_b - (U-R)_d \right> = 0.126$ mag, or $\left< (B-R)_b - (B-R)_d \right> = 0.045$ mag in an alternative filter combination.  More recently, \citet{mac04} reported a substantially larger mean bulge-disc color difference of $\left< (B-R)_b -(B-R)_d \right> = 0.29$ mag from an investigation of color gradients in a sample of 172 low-inclination, S0-Sd/Irr galaxies.  In that study component colors were estimated from the average colors at radii of 0-0.5$h$ for bulges and 1.5-2.5$h$ for discs.  \citet{mac04} speculated that their disagreement with Peletier \& Balcells' (1996) offset was either due to internal dust extinction, or a bias in the wedge-shaped aperture colors of the latter due to vertical gradients in stellar population age and metallicity within discs.  Here we recover an offset of $\left< (u-r)_b - (u-r)_d \right> = 0.22 \pm 0.02$ mag using our dust attenuation corrected \citep{dri08} colors, which is indeed also larger than Peletier \& Balcells' (1996) offset\footnote{We suppose a close agreement between $(u-r)$ and $(U-R)$ color differences based on the relevant filter transformation equations presented in Blanton \& Roweis' (2007) Table 2.}.  Removing the dust correction from our red peak systems results in an even larger offset for that population of $\left< (u-r)_b - (u-r)_d \right> = 0.27 \pm 0.04$ mag.

\section{Conclusions and Discussion}\label{discussion}
Our key conclusions may be summarized as follows.  Via bulge-disc decomposition and quantitative morphological classification (i.e., the logical filter in \citealt{all06}) of luminous ($M_B - 5 \log h_{70} < -18$ mag), nearby ($0.013 < z < 0.18$) galaxies in the MGC, we have demonstrated that: 

\textsc{(i)} one-component systems (predominantly elliptical and disc-only galaxies) display strong bimodality in the color-concentration plane, contributing to a red, high S\'ersic index peak and a blue, low S\'ersic index peak respectively (with total stellar mass densities of $0.7 \pm 0.1$ and $1.3 \pm 0.1$ $\times$ $10^8$ $h_{70}$ $M_\odot$ Mpc$^{-3}$ in each); 

\textsc{(ii)} luminous, `bulge-less, red discs' and `disc-less, blue bulges' (i.e., blue ellipticals) are rare; (with volume-densities of $1.7 \pm 0.3$ and $1.1 \pm 0.1$ $\times$ $10^{-4}$ $h^{3}_{70}$ Mpc$^{-3}$ respectively);

\textsc{(iii)} two-component systems (bulge-plus-disc galaxies) constitute a population of intermediate color-concentration, spanning the two peaks of the bimodality defined by one-component systems;

\textsc{(iv)} component bulge and disc colors are tightly correlated (with mean offset $\left<(u-r)_b-(u-r)_d\right> = 0.22 \pm 0.02$ mag); and

\textsc{(v)} the two-component systems may be subdivided in a physically-motivated manner at $(u-r)_g < 3.22-2.75 \log n$ into bridging (late-type disc) and red peak (early-type disc) populations (with total stellar mass densities of $1.1 \pm 0.1$ and $1.8 \pm 0.2$ $\times$ $10^8$ $h_{70}$ $M_\odot$ Mpc$^{-3}$ in each).

We now investigate how these results compare against other recent studies of galaxy structure, and discuss their interpretation within the hierarchical clustering scenario of galaxy formation, focussing on elliptical, disc-only, and bulge-plus-disc systems in turn.  Finally, we outline future work, and discuss the challenges facing future low resolution, automated structural decomposition campaigns.

\paragraph{Ellipticals}
The red colors and highly centrally-concentrated light profiles of luminous ellipticals in the MGC are consistent with many other observational studies, including a large sample of elliptical galaxies in the SDSS selected via a quantitative morphological approach and examined by \citet{ber03a,ber03b,ber03c,ber03d}.  Moreover, these properties are broadly consistent with the modern picture of hierarchical formation in which bright ellipticals are assembled at late epochs via `dry mergers' with other early-type systems \citep{kho03,del06}.  Only their progenitors (i.e., lower luminosity/mass ellipticals) are built up early via similar-sized disc mergers as in the original hierarchical model  \citep{too72,bar92}.  The modern scenario is supported by direct observations of early-type mergers at intermediate redshifts \citep{van99,tra05}, and N-body simulations confirming that `dry mergers' can reproduce the structural and kinematic properties of nearby ellipticals \citep{naa03} and preserve the Fundamental Plane \citep{nip03,boy05}. 

The `dry merger' scenario is also purported to account for the observed build up of mass on the red sequence since $z \sim 1$ \citep{bel04,bla06}.  However, studies showing minimal evolution of the early-type stellar mass-size relation \citep{mci05}---not preserved by `dry merging' \citep{nip03}---and constraints on the frequency of recent merging in luminous galaxies from pair counts and asymmetry indices \citep{dep07} challenge this hypothesis, suggesting a greater role for (non-merger) environmental processes in the recent evolution of the red sequence.

The study of galaxy structure via automated bulge-disc decomposition of large, intermediate-to-high redshift datasets, followed by comparison against local samples such as the MGC, may enable the degeneracy between early-type, bulge-plus-disc (S0) and predominantly elliptical (E) systems implicit in studies of the `red sequence' to be broken---thereby providing more accurate constraints on elliptical galaxy evolution.  However, the handling of `false disc' candidates (see Section \ref{false}) requires improvement.  In the meantime, the MGC elliptical population should serve as a key local benchmark, which semi-analytical models must be able to reproduce whilst simultaneously accounting for the properties of disc-only and bulge-plus-disc galaxies as well.

\paragraph{Disc-only systems}
The blue colors and diffuse (i.e., low-$n$) light profiles of predominantly disc-only systems in the MGC are also consistent with previous observations of disc-only galaxies, including a large sample of bulge-less, edge-on discs identified in the SDSS by \citet{kau06}.  These properties are also qualitatively consistent with expectations from the hierarchical formation scenario in which disc-dominated galaxies form early, but experience a passive evolutionary history \citep{fal80,mo98}.  However, the prevalence of disc-only systems in the local universe is difficult to reconcile with the expected merger history of galaxies in CDM cosmologies.  Specifically, in our analysis of the MGC data we identify disc-only systems as contributing $\sim$30\% of the total stellar mass in nearby disc galaxies, which is comparable to the number reported by \citet{kau06}.  In a theoretical study of halo merger histories in a $\lambda$CDM universe, \citet{kod07} reveal that the observed frequency of bulgeless discs can only be reconciled within the hierarhical formation context provided spheroid production only occurs in mergers involving mass ratios greater than 0.3.  Moreover, \citet{ste08} reveal equivalent limits on the frequency of minor mergers, which are expected to fuel central star-formation, leading to the creation of pseudobulges \citep{her95}.

One may question whether these disc-only galaxies identified in the MGC are indeed entirely `bulge-less', or whether they simply possess very faint `bulges', insignificantly brighter in central surface brightness than the discs in which they are embedded.  This is indeed possible given the limitations of our photometric decomposition pipeline \citep{all06}, and we further note the potential for dust extinction to obscure compact bulges \citep{tuf04,dri07b}.  Nevertheless, even if a population of very faint bulges is hiding within our disc-only population, their properties are vastly different to those of the bright, high surface brightness classical bulges predicted by the hierarhical clustering model \citep{bau96}.  We also note that in a high resolution study of 19 nearby late-type discs, \citet{boe03} observed that most of their sample could be well-fit using a single S\'ersic profile model, and the vast majority of central flux excesses detected could not be attributed to `what one would generally consider to be a bulge component'.  Similarly, \citet{wys97} point out that of the four largest disc galaxies in the local group, two (M33 and the LMC) do not possess convincing classical bulges.

However, it is certainly important to account for observational limitations when comparing output from semi-analytical simulations against empirical datasets in order to derive robust conclusions.  One example of how this may be achieved is the analysis pipeline developed by \citet{hau07} to generate realistic simulated galaxy images from model predictions, which may be used to compute `observed' structural parameters.

\paragraph{Bulge-plus-disc systems}
The close relationship between the bulge and disc colors of two-component galaxies---as well as recent low resolution \citep{gad08b,wei08} and high resolution studies \citep{dro07,lau07} revealing the disc-like nature of many blue \textit{and} red bulges---points towards a scenario in which \textit{both} secular evolution and merging play key roles in bulge formation, in contrast to the conventional hierarchical view \citep{bau96}.  Recent N-body simulations \citep{bou05,ath05,deb06} certainly indicate that secular evolutionary processes are efficient at establishing pseudobulge-like, central mass concentrations.  However, the distinction between classical bulge and pseudobulge systems may not be as clear as is often suggested---based upon SAURON observations of 24 Sa/Sab galaxies, \citet{pel08} report the co-existence of both an old, velocity-dispersion supported `classical' component and a young, rotationally-supported `pseudobulge' component within the nuclei of these early-type discs.

High redshift observations offer some further important clues as to the formation pathways of bulge-plus-disc galaxies, although cosmic variance contributes a large source of uncertainty to the results.  For instance, from the structural decomposition of 248 galaxies at $0.1 < z < 1.3$, \citet{dom08} reveal that the correlation betweeen bulge and global color is in place by intermediate redshifts.  They also identify 60\% of their bulges as displaying red colors consistent with passively-evolving stellar populations, and 40\% as displaying bluer colors indicative of more recent star-formation.  They speculate that the formation of red bulges coincided with a truncation of galaxy star-formation that did not destroy the disc.  Indeed, the role of black hole and AGN formation \citep{som08} cannot be overlooked in any complete model of bulge formation given the relationships between bulge and black hole properties \citep{mag98,fer00,geb00}.  In contrast to the results of \citet{dom08}, \citet{koo05} report a much higher fraction (85\%) of red bulges at high redshift in their sample of 71 cluster and 21 field galaxies at $z \sim 0.8$, more consistent with a merger-built origin.  However, the observed fraction is perhaps influenced by their selection of luminous bulges (rather than simply globally luminous galaxies).

\paragraph{Future work, and challenges for low resolution, automated structural decomposition studies}
Semi-analytical models of hierarchical galaxy formation (e.g.\ \citealt{van99b,col00}) now have the power to trace the formation and evolution of both bulge and disc components separately \citep{alm07}.  Moreover, wide-field galaxy surveys (e.g.\ SDSS, MGC, GEMS) now achieve sufficient survey volumes and imaging resolutions to facilitate large-scale, automated/semi-automated, bulge-disc decomposition studies \citep{tas05,all06,gad08b,wei08}.  A number of recent studies also offer important---albiet relatively small sample---benchmarks of galaxy structure in the intermediate-to-high redshift universe \citep{koo05,dom08}.  Hence, there currently exist unprecedented opportunities to unify theory and observation via detailed statistical studies of galaxies by structural type and component.

However, many observational challenges in this endeavor still remain.  Advanced, automated, data processing and quality control procedures are required to account for the impact of secondary structural features (e.g.\ bars and/or bright knots of active star-formation) during bulge-disc decompositions whilst ensuring meaningful output.  Automation of these processes is mandatory due to the vast sample sizes necessary to successfully limit cosmic variance uncertainties, provide accurate statistical constraints on the relevant structural parameters, and enable investigation of the role of environment and mass.   Finally, we note that a meaningful comparison between theory and observation will also require careful handling of the relevant observational selection biases and measurement errors, and the use of quantitative classification techniques for defining equivalent morphological sub-types.  Fortunately, quality, wide-field, near-IR survey data is becoming increasingly available (e.g.\ UKIDSS, GAMA), which should help to limit the impact of irregular star-formation features and dust attenuation.  Furthermore, the next generation of bulge-disc decomposition software is expected to introduce many new capabilities (as anticipated in the up-coming GALFIT v3.0).

\acknowledgments
The Millennium Galaxy Catalogue consists of imaging data from the Isaac Newton Telescope and spectroscopic data from the Anglo Australian Telescope, the ANU 2.3-m, the ESO New Technology Telescope, the Telescopio Nazionale Galileo and the Gemini North Telescope. The survey has been supported through grants from the Particle Physics and Astronomy Research Council (UK) and the Australian Research Council (AUS). The data and data products are publicly available from http://www.eso.org/\~jliske/mgc/ or on request from J.L.\ or S.P.D.  
E.C.\ acknowledges partial financial support from the Australian Research Council Discovery Project Grant DP0451426.

\begin{figure*}
\center
\epsscale{1.15}
\plotone{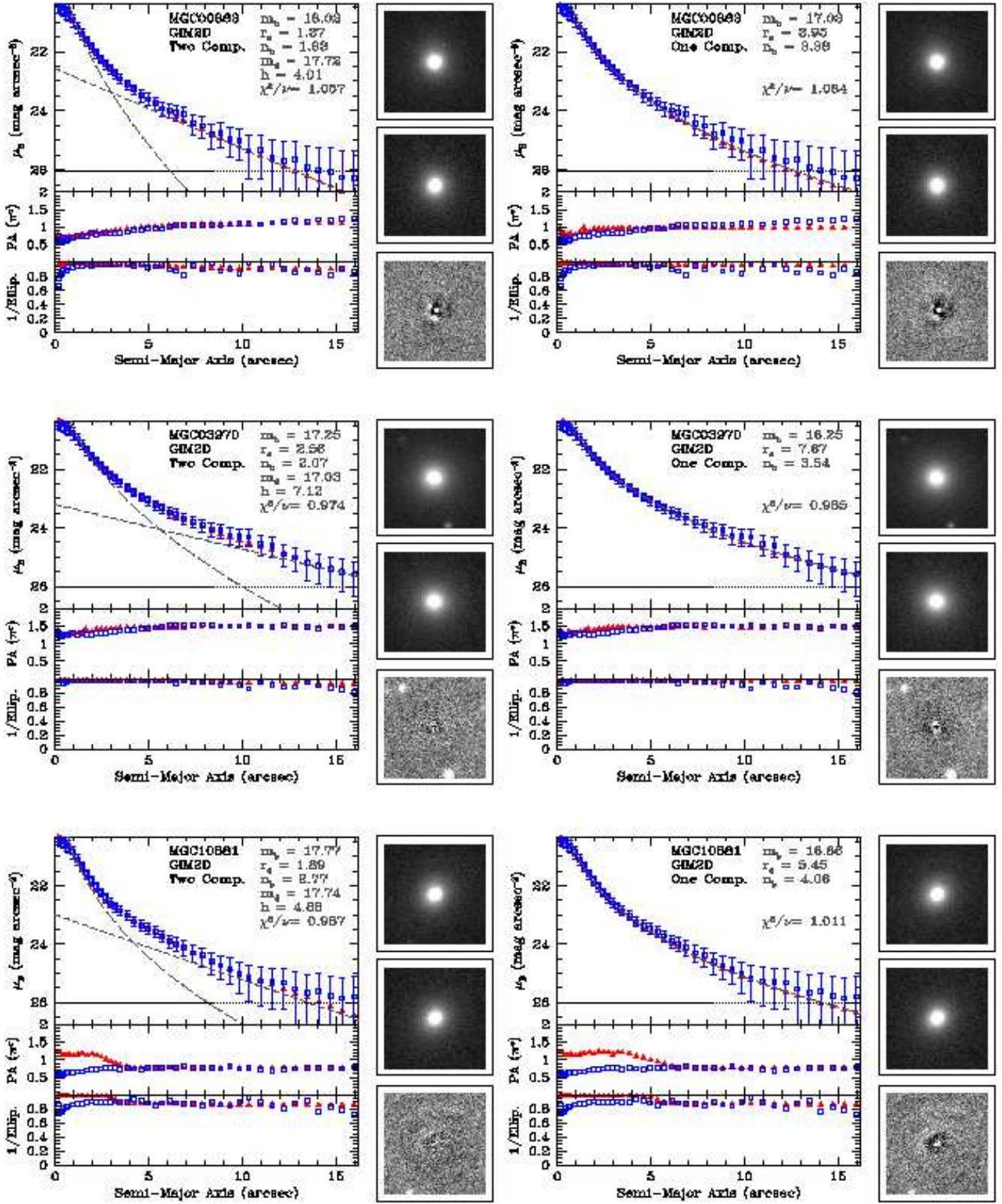}
\caption{\label{cut}Comparison of the GIM2D, two-component (S\'ersic bulge plus exponential disc) and one-component (S\'ersic-only) model fits to three bright MGC galaxies identified as two-component systems by the logical filter, but believed to be ellipticals based on visual inspection of their $B$-band images and surface brightness profiles.  In each case the elliptical isophote, semi-major axis, surface brightness profile of the real galaxy (recovered using the Starlink ELLPRO package) is indicated by open, blue squares and $3\sigma$ error bars.  The corresponding 2D model surface brightness profile (recovered from the PSF-convolved, GIM2D model output image, again using the ELLPRO package) is indicated by filled, red triangles, while the broken, black lines reveal the (non-PSF-convolved) contributions of each component separately.  Postage stamps of the $B$-band, MGC images, GIM2D model images and the corresponding residuals are also presented in greyscale with a logarithmic weighting function.  In each of these examples the addition of the outer exponential component results in only a trivial improvement over the one-component fit, as reflected in the relative reduced-$\chi^2$ values.  In the first example the improvement arises from fitting a mild asymmetry in the light distribution at intermediate radii, perhaps created in a recent merger event.  In the second and third examples the improvement arises from fitting a slight flux excess at intermediate-to-large radii, perhaps due to the presence of unresolved satellite(s).  As the two-component model is intended for `genuine' (i.e., rotationally-supported, flattened) exponential discs in this study, the S\'ersic-only fit was preferred for these suspected ellipticals, which were, therefore, returned to the one-component population.}
\end{figure*}

\begin{figure*}
\center
\epsscale{1.15}
\plotone{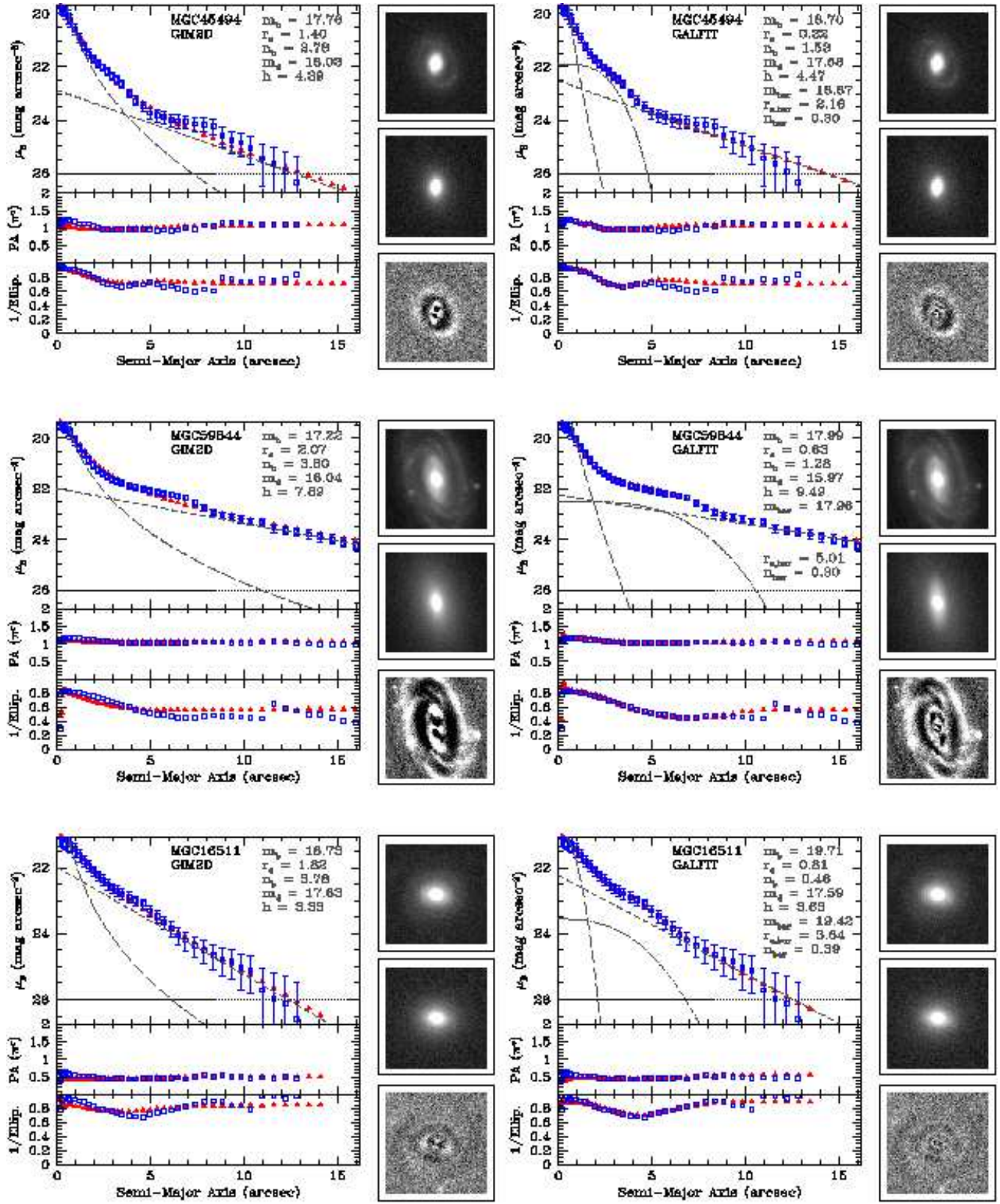}
\caption{\label{bars}Comparison of the GIM2D, two-component (S\'ersic bulge plus exponential disc) and GALFIT, three component (S\'ersic bulge plus S\'ersic bar plus exponential disc) model fits to three bright MGC galaxies observed to contain strong-to-intermediate bars upon visual inspection of their $B$-band images and surface brightness profiles.  In these examples the two-component model fails to provide a satisfactory representation of these intrinsically three (or more) component systems.  The presence of the bar results in systematic biases in the structural parameters recovered in the two-component fit.  Namely, a strong increase in the bulge S\'ersic index, half light radius, and flux, and a weak decrease in the disc scale-length and disc flux.  The labour-intensive option of refitting the MGC-BRIGHT sample with multi-component models using GALFIT or BUDDA will be explored in a future paper (Kelvin et al., in prep.).}
\end{figure*}

\begin{figure*}
\center
\epsscale{1.15}
\plotone{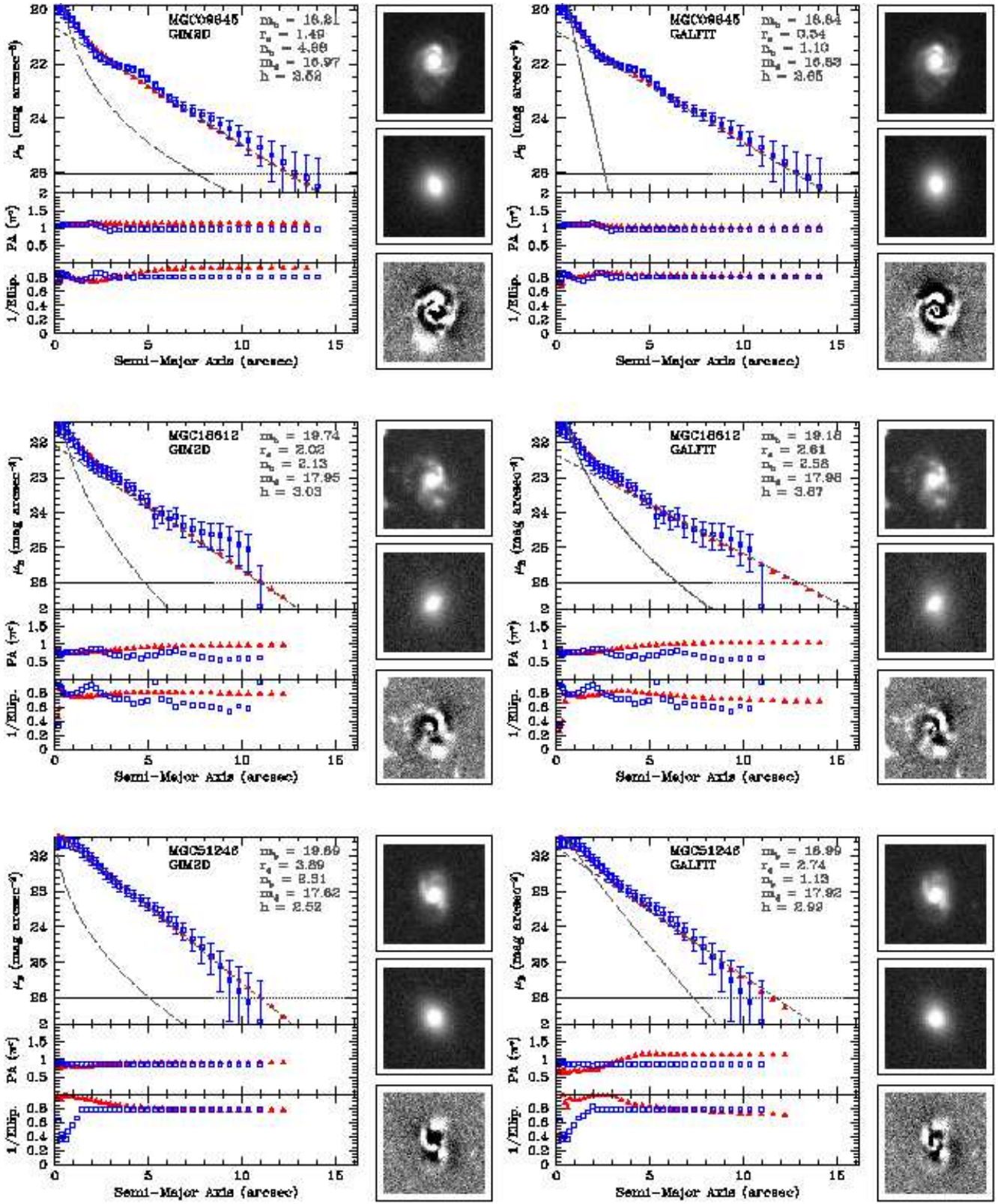}
\caption{\label{starform}Comparison of the GIM2D and GALFIT, two-component (S\'ersic bulge plus exponential disc) model fits to three bright MGC galaxies observed to be undergoing strong, localised, clumpy star-formation within the inner-to-intermediate disc region.  These examples illustrate the difficulty of recovering sensible bulge structural parameters from late-type disc galaxies, which typically possess only weak bulges and display much irregular structure within the disc.  Consequently, the most physically-meaningful, two-component fit may not always lie at a global, reduced-$\chi^2$ minimum, or it may be difficult to identify due to many nearby reduced-$\chi^2$ minima within the model parameter space.  Typically, this results in a systematic over-estimation of bulge S\'ersic index, half light radius, and flux.  In the first and third example GALFIT appears to have converged upon a better fit, and in the second example the GIM2D fit may be preferred.  Accounting for the effects of irregular patterns of star-formation in the disc remains a challenge for contemporary, structural decomposition software, expected to be addressed by the up-coming GALFIT version 3.0.}
\end{figure*}

\appendix
This Appendix contains real (MGC) and model (GIM2D/GALFIT), $B$-band images and surface brightness profiles of example galaxies illustrating three problematic fit scenarios---false discs (Fig.\ \ref{cut}), strong bars (Fig.\ \ref{bars}), and bright knots of active star-formation (Fig.\ \ref{starform}).

\end{document}